# Solution Blow Spinning Control of Morphology and Production Rate of Complex Superconducting YBa$_2$Cu$_3$O$_{7-x}$ Nanowires

M. Rotta[1,2], M. Motta[3], A. L. Pessoa[1], C. L. Carvalho[1], W. A. Ortiz[2], R. Zadorosny[1]

[1] Universidade Estadual Paulista (UNESP), Faculdade de Engenharia de Ilha Solteira, Av. Brasil 56, 15.385-000, centro, Ilha Solteira - SP, Brazil.
[2] Instituto Federal de Educação, Ciência e Tecnologia de Mato Grosso do Sul (IFMS), Campus Três Lagoas, Rua Ângelo Melão, 790, Jardim das Painciras, Três Lagoas/MS, 79641-162.
[3] Departamento de Física, Universidade Federal de São Carlos, UFSCar, SP, Brazil.

**Abstract**

The demand for nanostructured materials can increase exponentially due to the miniaturization of devices and their potential application in different areas, such as electronic and medicine. Therefore, high production rates are essential for making nanomaterials commercially available. When electrospinning (ES) and solution blow spinning (SBS) are employed for producing ceramic nanostructures, the solution injection rate can influence the morphology without, however, supply the real ceramic production rate. In this work, complex superconducting YBa$_2$Cu$_3$O$_{7-x}$ wires were prepared by using the SBS technique. We also show that the morphology can be controlled by varying the injection rate of the polymer solution and the production rate is 4.7 to 33 times higher than the rates of equivalent ceramics produced by ES. Additionally, we also suggest the term Ceramic Production Rate to refer to the production rate of ceramic structures.

# 1. Introduction

The remarkable properties of materials at the nanoscale have placed them into the focus of basic and applied research (1, 2, 3). To obtain nanomaterials, versatile and controllable techniques have been developed to tailor the intended properties of the specimens, as well as enhance their production rate (3, 4). Non-woven micro- and nanofibers have attracted a lot of attention since they can be applied as optical sensors and piezoelectric devices (5, 6, 7), once their sensitivity increases due to the larger surface area per unit mass. Among several methods to produce fibers (8), electrospinning (ES) (9, 10, 11, 12) and solution blow spinning (SBS) (11, 13, 14, 15, 16, 17) have recently gained prominence due to their simplicity and low cost. To produce ceramic fibers, ES and SBS can be combined with sol-gel processing to prepare nanowires for a variety of materials, such as $BaTiO_3$ (18, 19, 20), $TiO_2$ (14), ZnO (21), Mullite (22), and YBCO (17, 23).

The ES process has been known and used for more than a century (25) and has presented disadvantages such as a low production rate, requiring the application of high voltages, and the use of solutions with high dielectric constants (13). Alternatively, SBS is less expensive and does not require either solutions with high dielectric constants or a conducting collector. Additionally, its production rate for the final ceramic material can be higher than that of ES, making SBS a promising candidate for supplying an increased demand for nanofibers.

For both SBS and ES, the ratio between metallic salts (MS) and the amount of polymer (PM) is of great importance in the production and morphology of ceramic materials (13). However, several papers (11, 16, 17, 23) have not considered the ceramic production rate (CPR), reporting only the solution injection rate (SIR), which is a polymeric fiber productivity metric. However, as the same amount of precursor solution can have different MS:PM ratios (lower or higher quantities of metallic ions), the use of SIR as a measurement could create confusion about the productivity of ceramic. Thus, the adoption of CPR as a production metric would lead to more consistent and reliable production rate measurement, since it represents the production rate of the final ceramic fiber.

In this study, superconducting ceramic nanowires of $YBa_2Cu_3O_{7-\delta}$ were produced by SBS. We prepared samples of the precursor solution with two different values of SIR and, consequently, CPR. The crystalline structure and morphology of the samples were studied by X-ray diffraction (XRD) and scanning electron microscopy (SEM). Temperature-dependent magnetization curves were obtained in order to investigate the superconducting properties of the final specimens. Furthermore, parameters such as concentration of polymer and acetates, SIR, and CPR, are compared with data reported in other studies in literature using SBS and ES.

## 2. Experimental Procedures

In order to produce the precursor solution to be blown, stoichiometric amounts of acetates (Ac) of Y, Ba, and Cu were weighed in a molar ratio of 1:2:3. The amount of polyvinyl pyrrolidone (PVP) (MW 360,000 g/mol) was calculated based on a Ac:PVP weight ratio of 5:1. The final polymer solution was chosen to have a 5 wt% PVP concentration since in a previous analysis, such concentration resulted in a solution whose viscosity generated continuous jets and bead free fibrous samples. The acetates and PVP were dissolved in a solution of 14% propionic acid, 21% acetic acid, and 65% methanol (23).

Using the SBS technique, the precursor solution described previously was injected in an inner needle with two different values of SIR: (i) 60 μL/min (sample S6) and (ii) 100 μL/min (sample S10). The emerging jet was blown on a collector at a working distance of 40 cm with an angular velocity of 40 rpm by pressurized air released in an outer needle. In order to keep the polymer solution jet uniform (without interruptions and droplets), the airflow pressure was set to 66 ± 12 kPa and 50 ± 12 kPa for samples S6 and S10, respectively. It is important to emphasize that the main parameters that affect the morphology of the nanofibers produced by SBS are SIR, airflow pressure, working distance, and the viscosity of the solution. During this process the working distance was constant, the viscosities were equal, and the air flow pressure was consistent; the only adjustable parameter used to control the diameters of the nanofibers was the SIR (14, 26).

The collected fibers were calcined at 450°C for 3 hours to eliminate the organic precursors. The resultant materials were then given the following heat treatments in a flowing oxygen atmosphere: (i) 820°C for 14 hours, (ii) 925°C for 1 hour, (iii) 725°C for 3 hours, and (iv) 450°C for 12 hours (23). The influence of SIR on the microstructure of the samples was determined by using an EVO LS15 Zeiss SEM operated at 20 kV. XRD diffractograms were obtained in a Shimadzu XDR-6000 diffractometer with CuK$\alpha$ radiation. Measurements of the magnetization as a function of temperature were performed in a Quantum Design SQUID magnetometer MPMS-5S. The final ceramic fibers were measured under three different protocols: (i) Zero-Field Cooling (ZFC): the sample was cooled in a zero magnetic field (H=0) to the lowest temperature before H was applied, (ii) Field-Cooled Cooling (FCC): the sample was cooled to the superconducting critical temperature (Tc) under an applied magnetic field, and (iii) Thermoremanence (TRM): the specimen was cooled to the lowest temperature under H before then the magnetic field was set to zero and the measurement was done on warming.

## 3. Results and Discussion

Successful nanowire production using SBS is directly related to the synthesis of a precursor solution that provides high MS solubility in the polymer, resulting in a solution free of precipitates. The choice of acetates, the high solubility and volatility of the solvents, and the adoption of the PVP are of crucial importance. The PVP acts as a stabilizer, as its hydroxyl (-OH) limits metallic nanostructure production (27). Its high solubility enables the use of a high rate of 5:1 for Ac:PVP; this resulted in CPRs of 6.2 mg/min and 10.3 mg/min for samples S6 and S10, respectively (Table 1). By comparison, the ES approach yields CPR values which are roughly 3% of those obtained via SBS, meaning that SBS can generate a substantially higher production of ceramic nanowires.

Representative SEM micrographs and diameter distributions of the produced polymer fibers (green state), as well as the final ceramics, are shown in Figure 1. Panels (a) and (c) present the usual morphology of a cylindrical shape and smooth surface with no beads along the green fibers. Statistical analysis of fiber diameters returned an average value of 476 ± 92 nm for S6 (panel (b)) and 1222 ± 242 nm for S10 (panel (d)). This demonstrates that the average fiber diameter in the green state increases as the injection rate rises.

The SEM images in Figure 1 (e) and (g) show S6 and S10, respectively, after the heat treatments. The images demonstrate that the wires are randomly oriented and present a rough appearance due to their multigrain structure. The diameter distribution of S6 (Figure 1f) shows an average value of 258 nm and a standard deviation of 92 nm, which is 54% of the average of the green samples. S10 (Figure 1 (h)) presents an average diameter of 984 ± 204 nm, which is 80% of that of its green fiber. These results show that the elimination of the organic species results in ceramic wires with smaller diameters. Additionally, the diameters of S10 are nearly 4 times larger than those of S6, confirming the important role SIR plays in the diameter of SBS ceramic samples (13).

The final column in Table 1 compares the diameters of S6 and S10 with specimens prepared by ES. The diameter for S6 is similar to the sample obtained by Duarte et al. (17) with the same Ac:PVP ratio, although the production rate is 4.5 times larger. By using lower Ac:PVP ratios, Duarte et al. (17) and Shen et al. (28) created nanowires with smaller diameters than S6 and S10; however, the CPRs were also lower. It is important to point out that the samples prepared in reference 28 are YBCO nanotubes with wall thicknesses of 30-40 nm, which is a different morphology compared to our samples.

After all heat treatments and morphologic characterizations were completed, the presence of the YBaCuO superconducting phase, and also some secondary phases, was verified by XRD measurements. Figure 2 shows similar results for both samples, with no trace of spurious phases and consistent with the *JCPDS-78-2269* card. Using the Debye-Scherrer equation, crystallite sizes were estimated to be 20.7 ± 1.7 nm for S6 and 22.0 ± 1.4 nm for S10.

Considering the XRD patterns, a superconducting transition had been expected for both samples. Figure 3 presents magnetization versus temperature curves for samples S6 (panel (a)) and S10 (panel (b)). The superconducting critical temperature was determined by taking the last measured point in the normal state while decreasing the temperature. The $T_c$ values for S6 and S10 were 92.1 K and 93.1 K, respectively, similar to the value reported in references 27 and 28.

In superconducting specimens, the ZFC protocol is a procedure tailored to evaluate the amount of flux excluded from the sample, whereas the FCC is capable of sensing how much flux is expelled from the specimen. The difference between the ZFC and FCC curves is associated with the flux trapped in the material defects. Figure 3 (a) and (b) shows that there is more magnetic flux trapped in S10 than in S6; such behavior can be associated with a higher density of defects in the multigrain wires. Consistently, TRM curves confirm such behavior since they are a direct measurement of the flux trapped inside the samples.

## 4. Conclusion

SBS is a versatile and efficient technique for preparing complex ceramic wires. The average diameters of the YBCO fibers reached 258 nm and 984 nm with solution injection rates of 60 μL/min and 100 μL/min respectively, since small diameters increase the surface area, it was showed that SIR is an important parameter to obtain samples with the desirable sizes. Although SIR is the common parameter used to monitor the morphology of the fibers, the ceramic production rate is the most appropriate metric for such processes since it represents a reliable measurement of the real ceramic productivity. In our case, the CPR was 6.2 mg/min and 10.3 mg/min for samples S6 and S10, respectively, reaching values around 30 times greater than similar samples produced by the ES technique. Most importantly, the magnetic measurements confirmed that both samples, prepared under different conditions, exhibited a superconducting phase, with transition temperatures as high as 93 K.

## Acknowledgments

We acknowledge the Brazilian agencies São Paulo Research Foundation (FAPESP, grant 2016/12390-6), Coordenação de Aperfeiçoamento de Pessoal de Nível Superior - Brasil (CAPES) - Finance Code 001, and Instituto Federal do Mato Grosso do Sul (IFMS).

**Table 1:** Comparison of the solution injection rate (SIR), the ceramic production rate (CPR), and the diameter of the YBCO wire for the produced samples and data published in other studies.

| Sample | Ratio [Ac:PVP] | SIR (µL/min) | CPR (mg/min) | YBCO wire diameter range (nm) |
|---|---|---|---|---|
| **S6** | 5:1 | 60 | 6.2 | 206 – 310 |
| **S10** | 5:1 | 100 | 10.3 | 780 – 1188 |
| **Duarte et al. [17]** | 5:1 | 13.3 | 1.4 | 240 – 550 |
| **Duarte et al. [17]** | 1:1 | 13.3 | 0.3 | 120 – 300 |
| **Shen et al. [28]** | 1:1.6 | 50 | 2.2 | 110 – 150* |

* This range refers to the external diameter of YBCO nanotubes with a wall thickness of 30 – 40 nm.

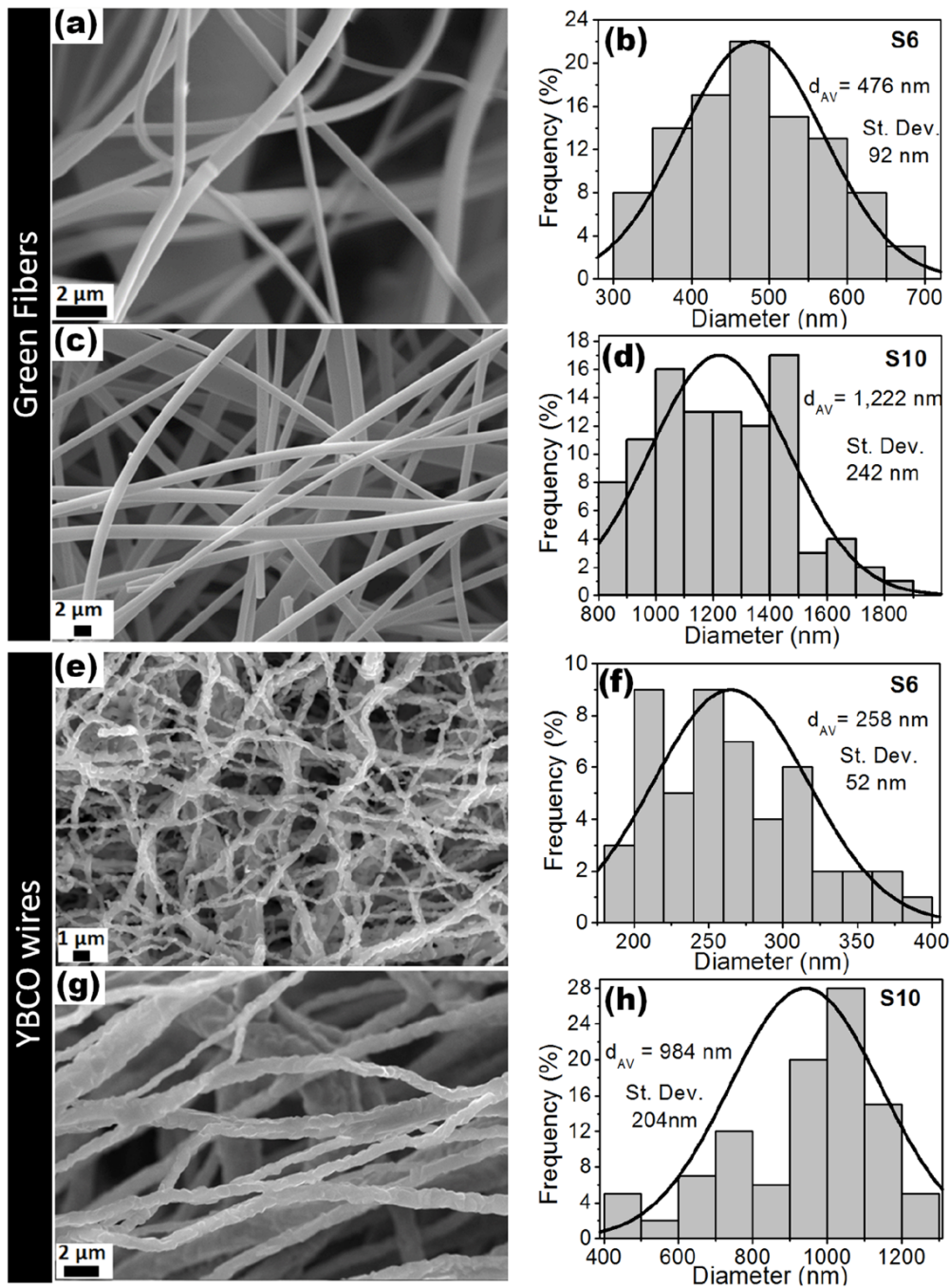

**Figure 1:** SEM micrograph of sample S6 (a) and its diameter distribution with an average value of 476 ± 92 nm (b). Panel (c) shows a SEM micrograph of sample S10 in green-state with an average diameter of 1222 ± 242 nm (d). The SEM images of the fired fibers S6 and S10 and their diameter distributions are shown in panels (e) through (h). The average diameter values were 258 ± 52 nm for S6 (f) and 984 ± 204 nm for S10 (h).

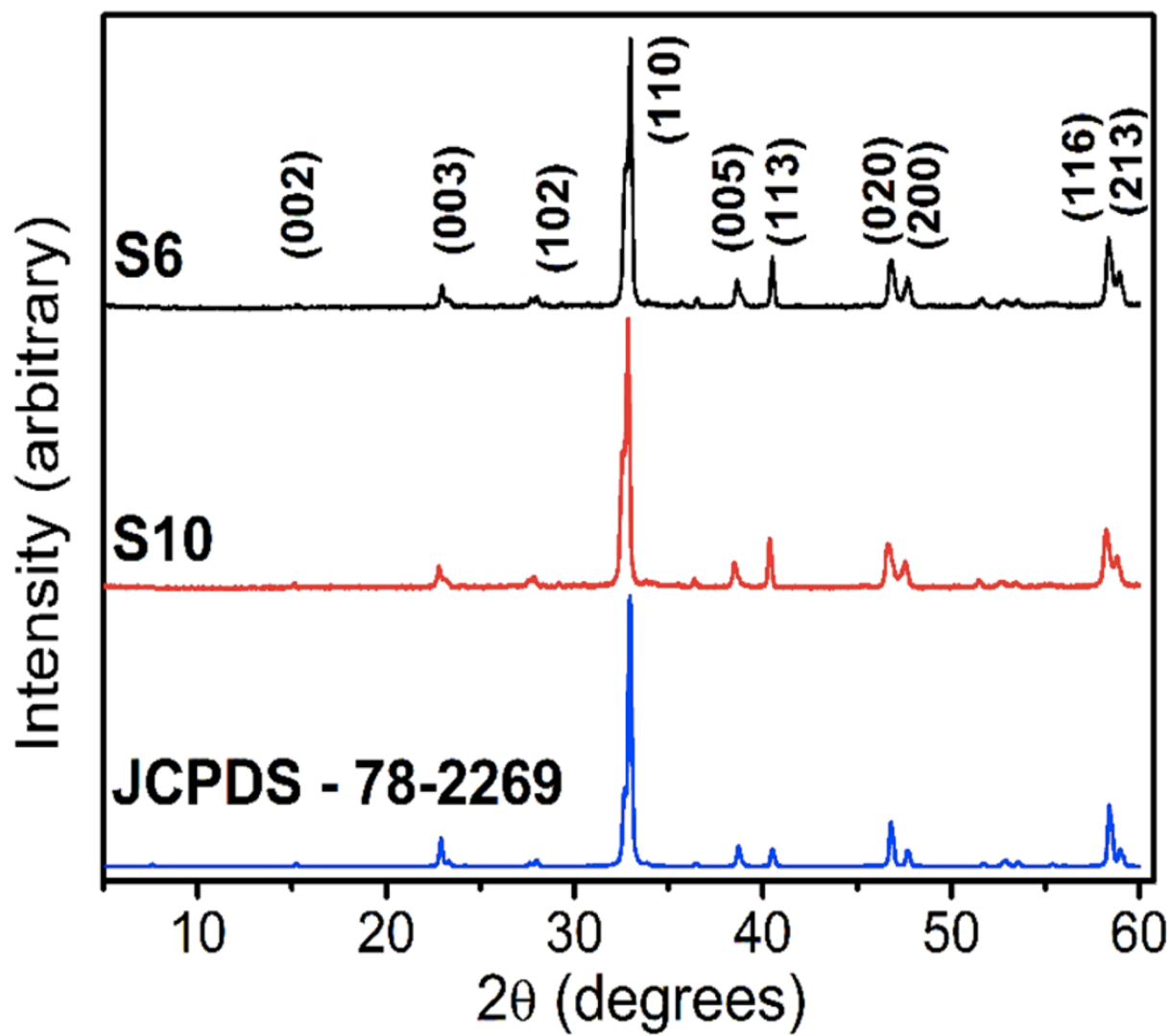

**Figure 2:** XRD diffractograms taken at a scan rate of 0.02°/min for samples S6 and S10, which are consistent with the *JCPDS-78-2269* reference card.

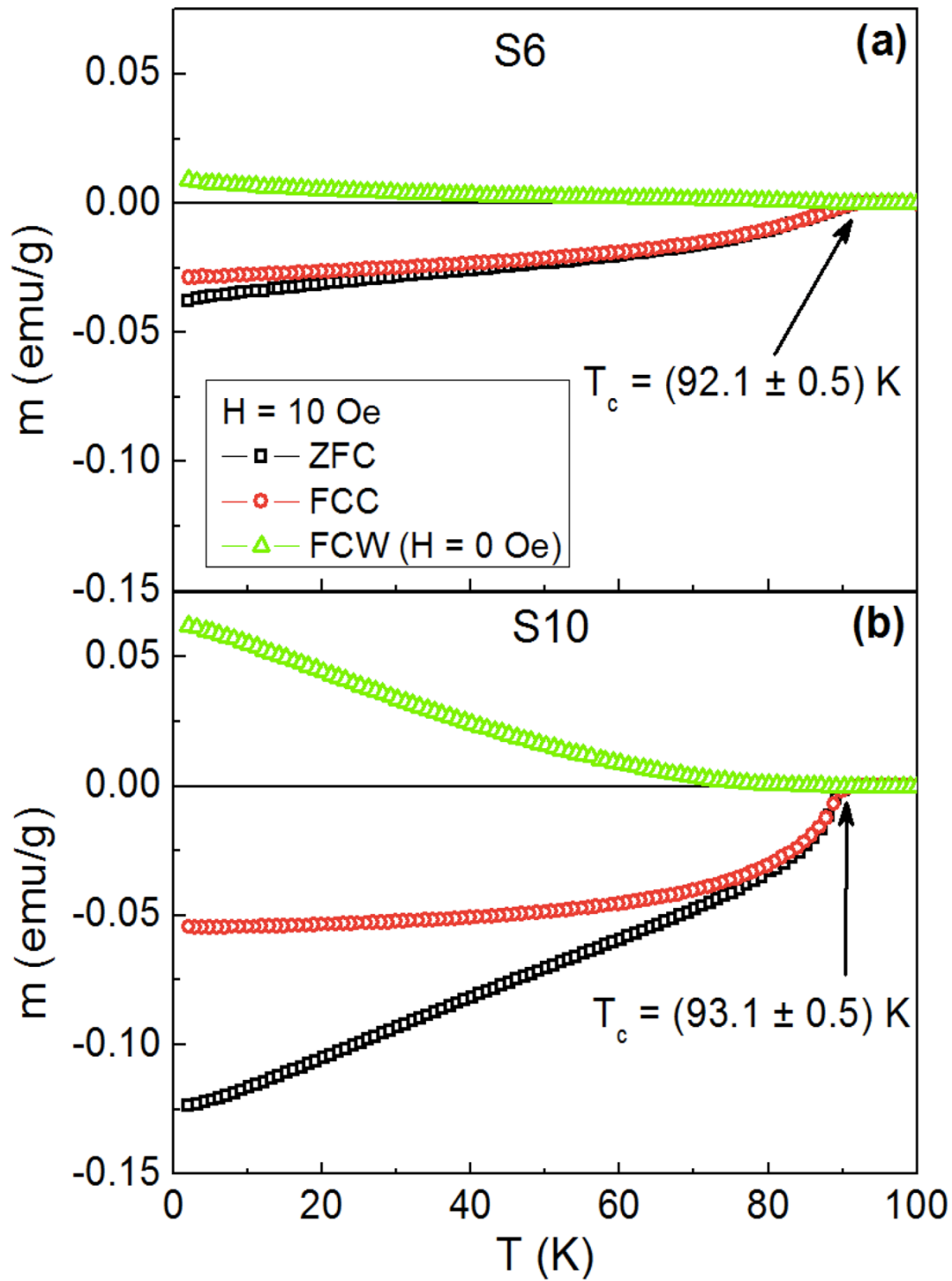

**Figure 3:** Measurements of the magnetization as a function of temperature following the procedures ZFC, FCC, and TRM (see the main text) for samples S6 (a) and S10 (b), taken under an applied field H = 10 Oe with $T_c$ of 92.1 K and 93.1 K, respectively.